\documentclass[sn-basic,iicol]{sn-jnl}

\jyear{2022}%

\theoremstyle{thmstyleone}%
\theoremstyle{thmstyletwo}%

\theoremstyle{thmstylethree}%

\newcommand{\viecrfname}{VIE-K-2022b}
\newcommand{\usnocrfname}{USNO-K-2022July05}
\begin{document}

\title[Article Title]{The K-band (24 GHz) Celestial Reference Frame determined from Very Long Baseline Interferometry sessions conducted  over the past 20~years}


\author*[1]{\fnm{Hana} \sur{Kr\'asn\'a}}\email{hana.krasna@tuwien.ac.at}

\author[2]{\fnm{David} \sur{Gordon}}

\author[3]{\fnm{Aletha} \sur{de Witt}}

\author[4]{\fnm{Christopher S.} \sur{Jacobs}}

\affil*[1]{\orgdiv{Department of Geodesy and Geoinformation}, \orgname{Technische Universit\"at
Wien}, \orgaddress{\street{Wiedner Hauptstraße 8-10/E120.4}, \city{Vienna}, \postcode{1040}, \country{Austria}}}

\affil[2]{\orgname{United States Naval Observatory}, 
\orgaddress{ \country{USA}}}

\affil[3]{\orgname{South African Radio Astronomy Observatory}, \orgaddress{\country{South Africa}}}

\affil[4]{\orgname{Jet Propulsion Laboratory, California Institute of Technology}, 
\orgaddress{\country{USA}}}


\abstract{The third realization of the International Celestial Reference Frame (ICRF3) was adopted in August 2018 and includes positions of extragalactic objects at three frequencies: 8.4~GHz, 24~GHz, and 32~GHz. In this paper, we present celestial reference frames estimated from Very Long Baseline Interferometry measurements at K-band (24 GHz) including data until June 2022. The data set starts in May 2002 and currently consists of more than 120 24h observing sessions performed over the past 20 years. Since the publication of ICRF3, the additional observations of the sources during the last four years allow maintenance of the celestial reference frame and more than 200 additional radio sources ensure an expansion of the frame. A study of the presented solutions is carried out helping us to understand systematic differences between the astrometric catalogs and moving us towards a better next ICRF solution. We compare K-band solutions (\viecrfname{} and  \usnocrfname{}) computed by two analysts with two independent software packages (VieVS and Calc/Solve) and describe the differences in the solution strategy. We assess the systematic differences using vector spherical harmonics and describe the reasons for the most prominent ones.}




\maketitle

\section{Introduction}\label{intro}
The current International Celestial Reference Frame \citep [ICRF3;][]{Charlot20} is the third realization of the International Celestial Reference System adopted by the International Astronomical Union in August 2018. The ICRF3 is the first multi-wavelength radio frame since it contains positions of active galactic nuclei (AGN) observed with Very Long Baseline Interferometry (VLBI) at 2.3 and 8.4~GHz (S/X-band), 24~GHz (K-band), and 8.4 and 32~GHz (X/Ka-band). The three components differ as shown by several statistical indicators (e.g., data span, number of sources, coordinate uncertainty, error ellipse) and each of them faces different challenges. In 2018 IAU Resolution B2, ``On The Third Realization of the International Celestial Reference Frame," \citep{iau18} recommended that appropriate measures should be taken to both maintain and improve ICRF3. In response, this paper concentrates on the two main challenges in improving the accuracy of the celestial reference frame observed at K-band (K-CRF) which are (1) observations at a single frequency requiring an external ionospheric calibration and (2) the lack of a uniform global terrestrial network causing a non-optimal observation geometry. Our main goal is to assess systematic differences in the K-CRF solutions which are computed at two VLBI analysis centers: at TU Wien with VLBI software package VieVS~\citep{Boehm18} and at the United States Naval Observatory (USNO) with Calc/Solve. We also compare these two frames to the ICRF3 using vector spherical harmonics (VSH) which provides information about systematic differences between pairs of astrometric catalogs and we investigate the possible reasons for the estimated differences.

\section{Data and solution setup}\label{data}
\subsection{Data description}

\begin{table*}
\caption{Overview of sessions included in our solutions listed with recording rate.}
\label{tab:ses}%
\begin{tabular}{llr}
\toprule
time span & session code & data rate [Mbps] \\
\midrule
& Northern (VLBA) sessions  &   \\
05/2002 - 12/2008    & BR079a-c, BL115a-c, BL122a-d, BL151a-b   & 128   \\
06/2006 - 10/2006    & BP125a-c   &  256    \\
12/2015 - 10/2019    & BJ083a-d, UD001a-x, UD009a-o   & 2048   \\
11/2019 - 06/2022    & UD009p-z, UD009aa-ah, UD015a-l   &  4096   \\
\midrule
& Southern sessions   &   \\
05/2014 - 07/2016    & KS1401, KS1601   &  1024   \\
11/2016 - 02/2021    & KS1603, KS1702-KS2102   &  2048\\
\botrule
\end{tabular}
\end{table*}

\begin{table*}
\scrisize
\caption{Selected models and parametrization in \viecrfname{} and  \usnocrfname{}. Values in parentheses represent the applied constraints. The abbreviation pwlo stands for piecewise linear offset.}
\label{tab:statmodel}       
\begin{tabular}{lll}
\toprule
 A priori modeling& \viecrfname{} & \usnocrfname{} \\
\midrule
ionosphere maps & CODE time series \citep{Schaer1999} & 2 h average JPL maps\\
ionospheric mf coefficients& MSLM, $k$ = 1, $\Delta H$ = 56.7~km, $\alpha$ = 0.9782 & 2-D thin shell, MSLM \\
hydrostatic delay& in situ pressure \citep{Saastamoinen1972} & in situ pressure \citep{Saastamoinen1972}\\
hydrostatic + wet mf & VMF3 \citep{Landskron2018} & VMF1 \citep{Boehm2006}\\
hydrostatic gradients &  DAO \citep{MacMillan1997} & DAO \citep{MacMillan1997} \\
precession/nutation model & IAU 2006/2000A  & IAU 2006/2000A\\
celestial pole offsets (CPO)& IERS Bulletin A, \href{finals2000A.all}{https://maia.usno.navy.mil/ser7/finals2000A.all} & none\\
\midrule
Parametrization& &\\
\midrule
zenith wet delay & 30 min pwlo (1.5~cm/30 min) & 30 min pwlo (1.5 cm/h)\\
tropo. grad.: VLBA & 3 h pwlo (0.5~mm/3 h) & 6 h pwlo (0.5mm, 2 mm/day)\\
tropo. grad.: KS & fixed to a priori & fixed to a priori\\
CPO: VLBA & 24 h pwlo (0.1~\textmu as/24 h) & offset at midpoint of the session\\
CPO: KS & fixed to a priori & fixed to a priori \\
weighting & elevation-dependent \citep{Gipson08}& baseline-dependent\\
\botrule
\end{tabular}
\end{table*}

The celestial reference frames introduced in this paper are computed from $1.96\cdot10^6$ group delays observed at K-band in the VLBI sessions listed in Table~\ref{tab:ses}. This data set was acquired mainly with the Very Long Baseline Array (VLBA) starting in May 2002. The first sessions belong to programs carried out by \cite{Lanyi2010} and \cite{Petrov2011}. All sessions up to May 2018 are part of the current ICRF at K-band, ICRF3-K. The VLBA \citep{napier95}, because its sites are limited to U.S. territory, does not allow observations of sources with declinations below $-46^\circ$. Therefore, southern K-band sessions (KS) were organized starting in May 2014. The vast majority of southern observations are from single baseline sessions between the HartRAO 26m (South Africa) and the Hobart 26m (Tasmania, Australia) with the exception of one session involving the Tianma 65m (near Shanghai, China) and four sessions augmented with the Tidbinbilla 70m telescope (near Canberra, Australia). Of all the sources, 913 were observed in VLBA sessions, 328 were observed in southern hemisphere sessions, and 206 were observed in both types between $-46^\circ$ and $+39^\circ$ declination.

\begin{figure}[h]%
\centering
\includegraphics[width=0.5\textwidth]{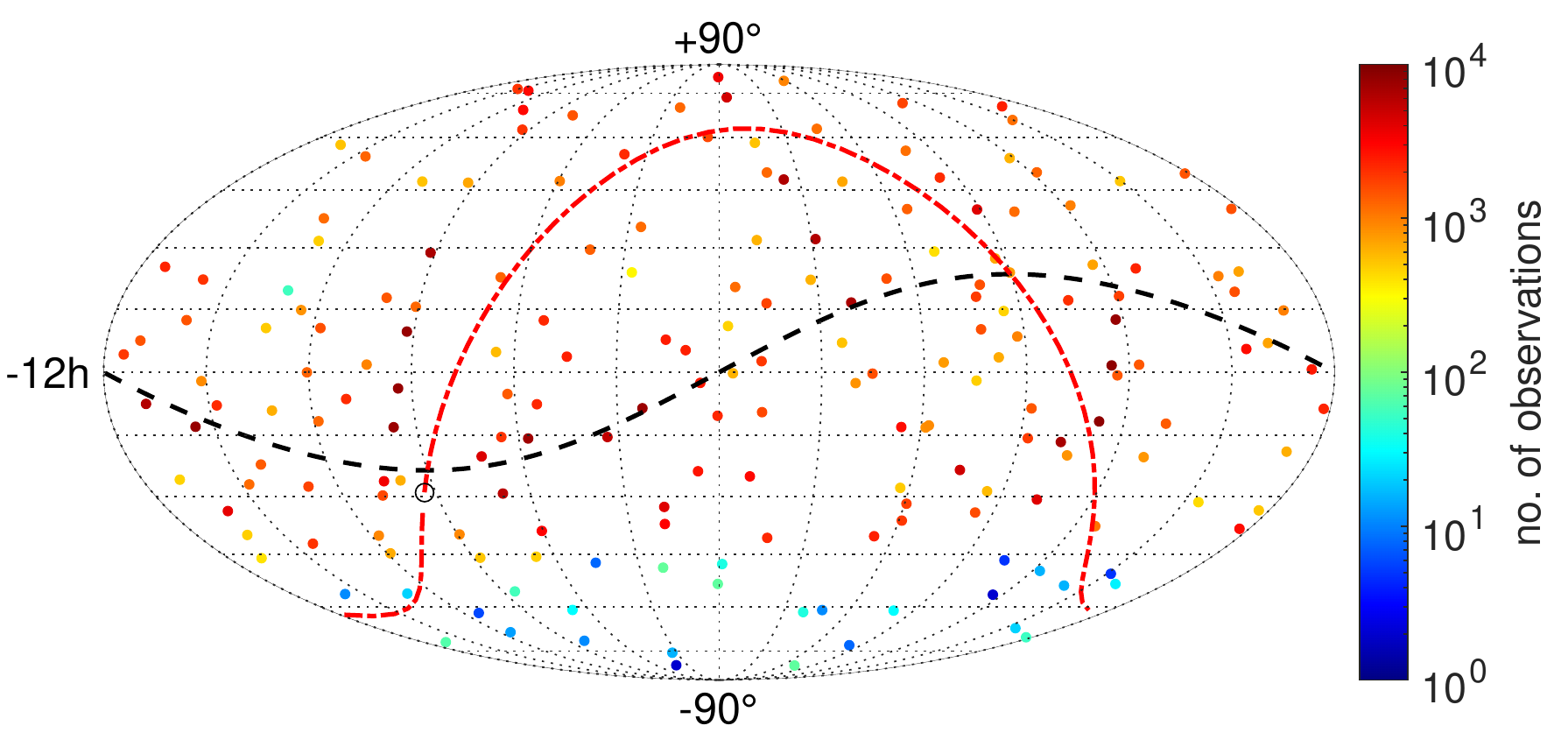}
\includegraphics[width=0.5\textwidth]{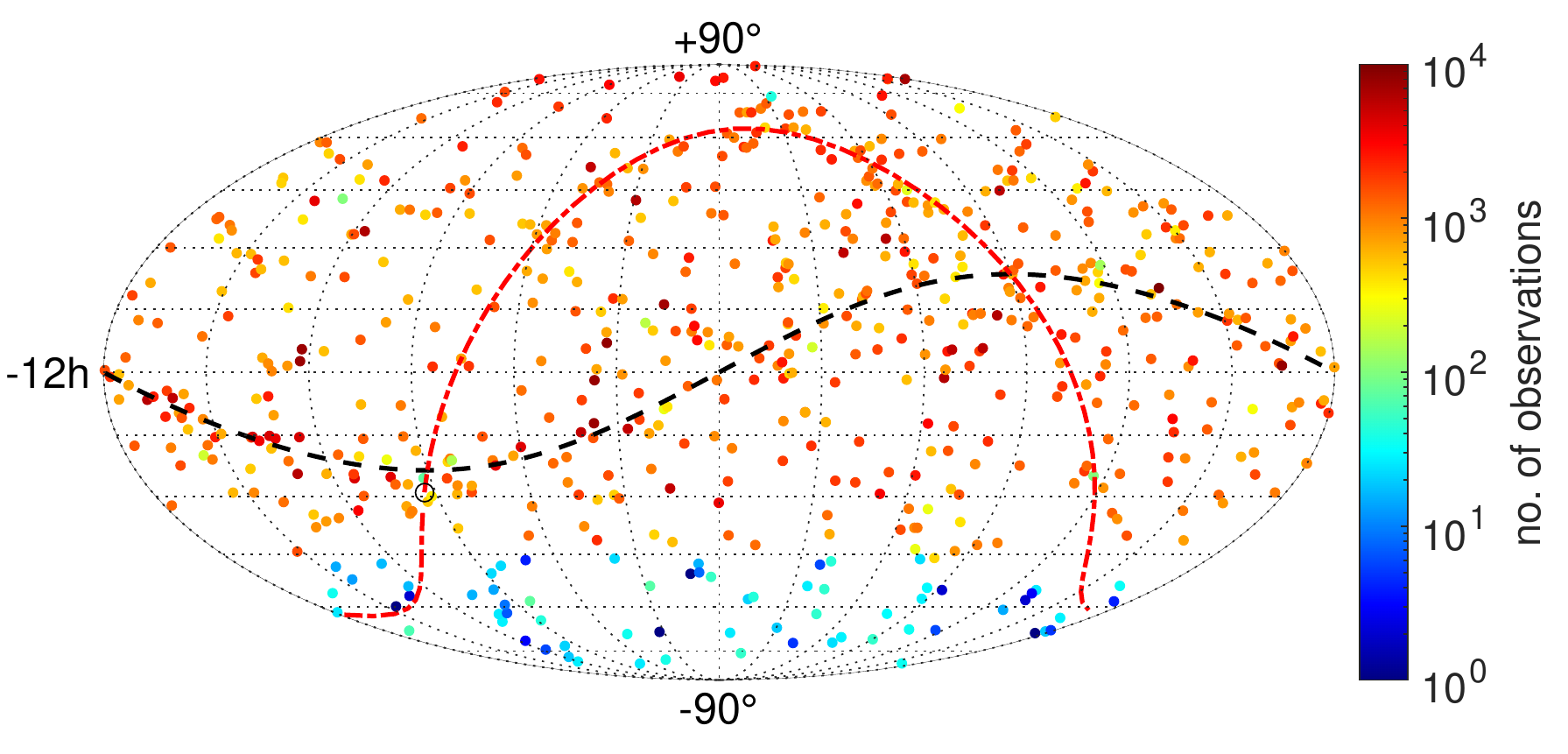}
\includegraphics[width=0.5\textwidth]{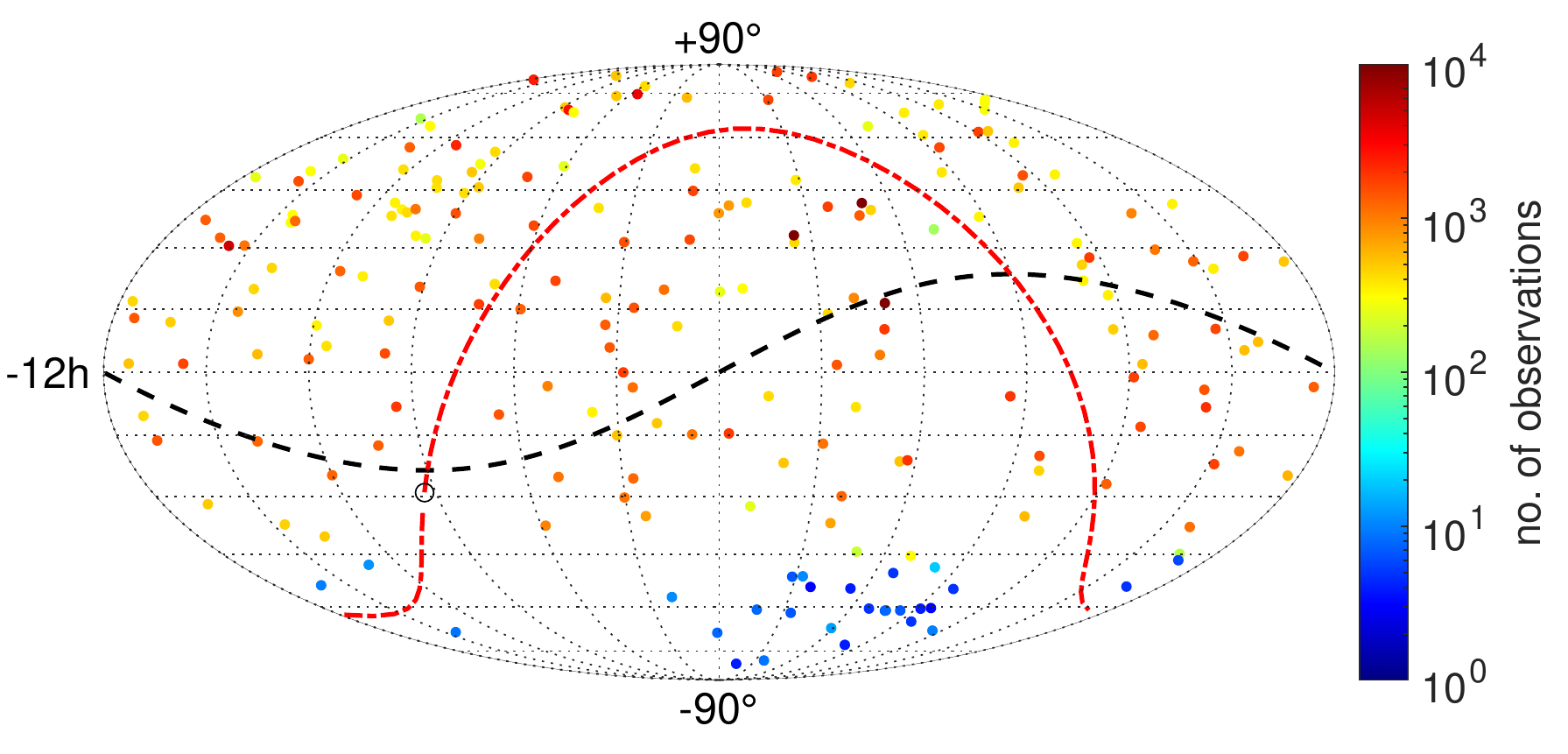}
\caption{Number of observations after the ICRF3-K data cutoff in May 2018 until June 2022. The sources are divided into three groups: ICRF3-K defining sources (top), ICRF3-K non-defining sources (middle), and sources not included in ICRF3-K (bottom).}\label{fig_numobs_wrtICRF3}
\end{figure}

In Fig.~\ref{fig_numobs_wrtICRF3} we show the number of observations conducted after the ICRF3 K-band data cutoff on 5 May 2018 until June 2022 divided into three groups: (a) observations to ICRF3-K defining sources, (b) observations to ICRF3-K non-defining sources and (c) observations to sources which are not included in ICRF3-K. The consequence of using mainly the VLBA network for the K-band observations is the lack of observations of the deep south sources which is currently amplified by the technical problems of Hobart26 since March 2021. The low number of new observations (under 100) of the deep south sources since the ICRF3 release is seen in all three plots of Fig.~\ref{fig_numobs_wrtICRF3}.

\subsection{Setup of solutions}

The treatment of the K-band VLBI observations in the VieVS solution (\viecrfname{}) is similar to the S/X solution VIE2022b computed at the VIE Analysis Center\footnote{\href{https://www.vlbi.at}{https://www.vlbi.at}} of the International VLBI Service for Geodesy \& Astrometry. A detailed description of the setup and applied theoretical models during the analysis are given in \cite{Krasna2022}. In Table~\ref{tab:statmodel} we highlight models used in \viecrfname{} and the USNO Calc/Solve solution \usnocrfname{}\footnote{latest version at \href{https://crf.usno.navy.mil/data\_products/RORFD/Quarterly/current//USNO_Kband_source_positions.iers}{https://crf.usno.navy.mil/data\_products/RORFD/Quarterly/current//USNO_Kband_source_positions.iers}} 
relevant to the presented investigations. While S/X frames calibrate the ionosphere directly from their dual-band data, K-band ionospheric effects require external calibration data. Specifically, K-band systems at the VLBA and the southern stations currently lack the complementary lower band needed for a dual-band ionospheric calibration, therefore the frequency-dependent delay coming from the dispersive part of the atmosphere has to be described by external models. In both K-band solutions presented here, ionospheric maps derived from Global Navigation Satellite System (GNSS) are applied. In \viecrfname{}, global ionospheric maps provided by the Center for Orbit Determination in Europe \citep [CODE;][]{Schaer1999}\footnote{\href{http://ftp.aiub.unibe.ch/CODE/}{http://ftp.aiub.unibe.ch/CODE/}} are used with a time spacing of two hours from 05/2002 until 05/2014, and of one hour since that date. In \usnocrfname{}, global ionospheric maps computed at the Jet Propulsion Laboratory (JPL) with two hours resolution are applied.\\
The alignment of the Terrestrial Reference Frame (TRF) is done by applying the No-Net-Translation (NNT) and No-Net-Rotation (NNR) conditions to the station position and velocity parameters in the global normal matrix. In \viecrfname{}, the conditions are applied to all VLBA telescopes but one (MK-VLBA) with respect to the ITRF2020. In \usnocrfname{}, the NNT/NNR condition is used w.r.t. a TRF solution based on ITRF2014 applied to all participating antennas except MK-VLBA (position discontinuity due to an Earthquake on June 15, 2006) and TIDBIN64 (limited number of observations). 

The common practice for the rotational alignment of a new celestial reference frame to the current official one is to apply a three-dimensional constraint to the defining sources. In both solutions, ICRF3-SX is used as a priori celestial reference frame and the galactic acceleration correction is modeled with the adopted ICRF3 value of 5.8~\textmu as/yr for the amplitude of the solar system barycenter acceleration vector for the epoch 2015.0. Datum definition of the CRFs is accomplished by the unweighted NNR \citep{Jacobs2010} w.r.t. 287 (\viecrfname{}) and 258 (\usnocrfname{}) defining ICRF3-SX sources.

\section{Results}\label{res}

We analyze the estimated \viecrfname{} and \usnocrfname{} frames in terms of the 
vector spherical harmonics decomposition \citep[VSH; ][]{Mignard2012, Titov2013, mayer2020} w.r.t. ICRF3-SX which allows studying possible systematic differences between the catalogs. Prior to the comparison, outliers---defined as AGN with an angular separation greater than 5~mas from their ICRF3-SX position---were removed. In both solutions, there are four outlier sources: 
0134+329 (3C48),
0316+162 (CTA21),
0429+415 (3C119), and
2018+295. Note that large position changes for 3C48 and CTA21 were found at X-band in observations made after the ICRF3 release and are reported by \citet{Frey2021} and \citet{Titov2022}.
The number of remaining common sources is 993 in \viecrfname{} and 995 in \usnocrfname{}. The two sources (0227-542 and 0517-726) missing in \viecrfname{} have 3 and 4 observations in \usnocrfname{}. In \viecrfname{} these observations were removed based on an outlier check of individual observations during the single session analysis. 

The VSH are obtained with a least squares adjustment where the weight matrix contains inflated formal errors of the source coordinates. Similar to ICRF3-K, the formal errors of the source coordinates in both catalogs are inflated by a factor of 1.5, and a noise floor of 30 and 50~\textmu as in quadrature is added to right ascension and declination, respectively. Table~\ref{tab:souVSH} summarizes the first order and second degree and order VSH, i.e., rotation ($R_1, R_2, R_3$), dipole ($D_1, D_2, D_3$), and ten coefficients ($a$) for the quadrupole harmonics of magnetic ($m$) and electric ($e$) type. All three rotation angles between the \viecrfname{} and ICRF3-SX axes are within their formal errors and the angles do not exceed 8~\textmu as. The largest angle ($16 \pm 10$~\textmu as) between \usnocrfname{} and ICRF-SX is around the y-axis ($R_2$). The selection of defining sources for the NNR constraint influences the mutual rotations of two catalogs (cf. Section~\ref{sec_defsou} for more details). The three dipole parameters represent the distortion as a flow from a source to a sink located at two opposite poles. The $D_3$ term ($-4 \pm 10$~\textmu as in \viecrfname{} and $60 \pm 9$~\textmu as in \usnocrfname{}) is susceptible to imperfect modeling of equatorial bulges in the ionospheric and tropospheric calibrations (cf. Section~\ref{sec_iono}). The zonal quadrupole terms $a_{2,0}^{e}$ and $a_{2,0}^{m}$ reflect north-south asymmetries. Their values w.r.t. ICRF3-SX reach $-3 \pm 12$~\textmu as and $-36 \pm 7$~\textmu as in \viecrfname{}, and $-46 \pm 11$~\textmu as and $1 \pm 7$~\textmu as in	\usnocrfname{}, respectively (cf. Section~\ref{sec_syselev}).

\begin{table}
\caption{VSH parameters up to degree and order two for \viecrfname{} and \usnocrfname{} w.r.t. ICRF3-SX (after eliminating four outliers from the solutions).}
\label{tab:souVSH}       
\begin{tabular}{l|r|r}
\hline\noalign{\smallskip}
[\textmu as] & \viecrfname{}& \usnocrfname{}\\
\noalign{\smallskip}
\hline 
$R_1$ &  $ -1 ~\pm 10$ 		&  $  -4 ~\pm 10$ \\ 
$R_2$ &  $ -8 ~\pm 10$ 		&  $ -16 ~\pm 10$ \\ 
$R_3$ &  $ +0 ~\pm ~6$ 		&  $ -11 ~\pm ~6$ \\ 
\hline 
$D_1$ &  $-17 ~\pm ~9$ 		&  $ -5 ~\pm ~9$ \\ 
$D_2$ &  $-15 ~\pm ~9$ 		&  $ +9 ~\pm ~9$ \\ 
$D_3$ &  $ -4 ~\pm 10$ 		&  $+60 ~\pm ~9$ \\ 
\hline 
$a_{2,0}^{e}$    & $ -3 ~\pm 12$ 	& $-46 ~\pm 11$ 	 \\
$a_{2,0}^{m}$    & $-36 ~\pm ~7$    & $ +1 ~\pm ~7$ 	 \\
$a_{2,1}^{e,Re}$ & $-19 ~\pm 10$ 	& $-13 ~\pm 10$ 	 \\
$a_{2,1}^{e,Im}$ & $-21 ~\pm 11$ 	& $-26 ~\pm 11$ 	 \\
$a_{2,1}^{m,Re}$ & $-13 ~\pm 11$ 	& $+13 ~\pm 10$ 	 \\
$a_{2,1}^{m,Im}$ & $-12 ~\pm 11$ 	& $ -6 ~\pm 11$ 	 \\
$a_{2,2}^{e,Re}$ & $ +1 ~\pm ~4$ 	& $ +3 ~\pm ~4$ 	 \\
$a_{2,2}^{e,Im}$ & $ +8 ~\pm ~4$ 	& $ +3 ~\pm ~4$ 	 \\
$a_{2,2}^{m,Re}$ & $+12 ~\pm ~5$ 	& $+23 ~\pm ~5$ 	 \\
$a_{2,2}^{m,Im}$ & $ +6 ~\pm ~5$ 	& $ +4 ~\pm ~5$ 	 \\
\noalign{\smallskip}\hline
\end{tabular}
\end{table}

\subsection{Defining sources}
\label{sec_defsou}

\begin{figure}[h]%
\centering
\includegraphics[width=0.5\textwidth]{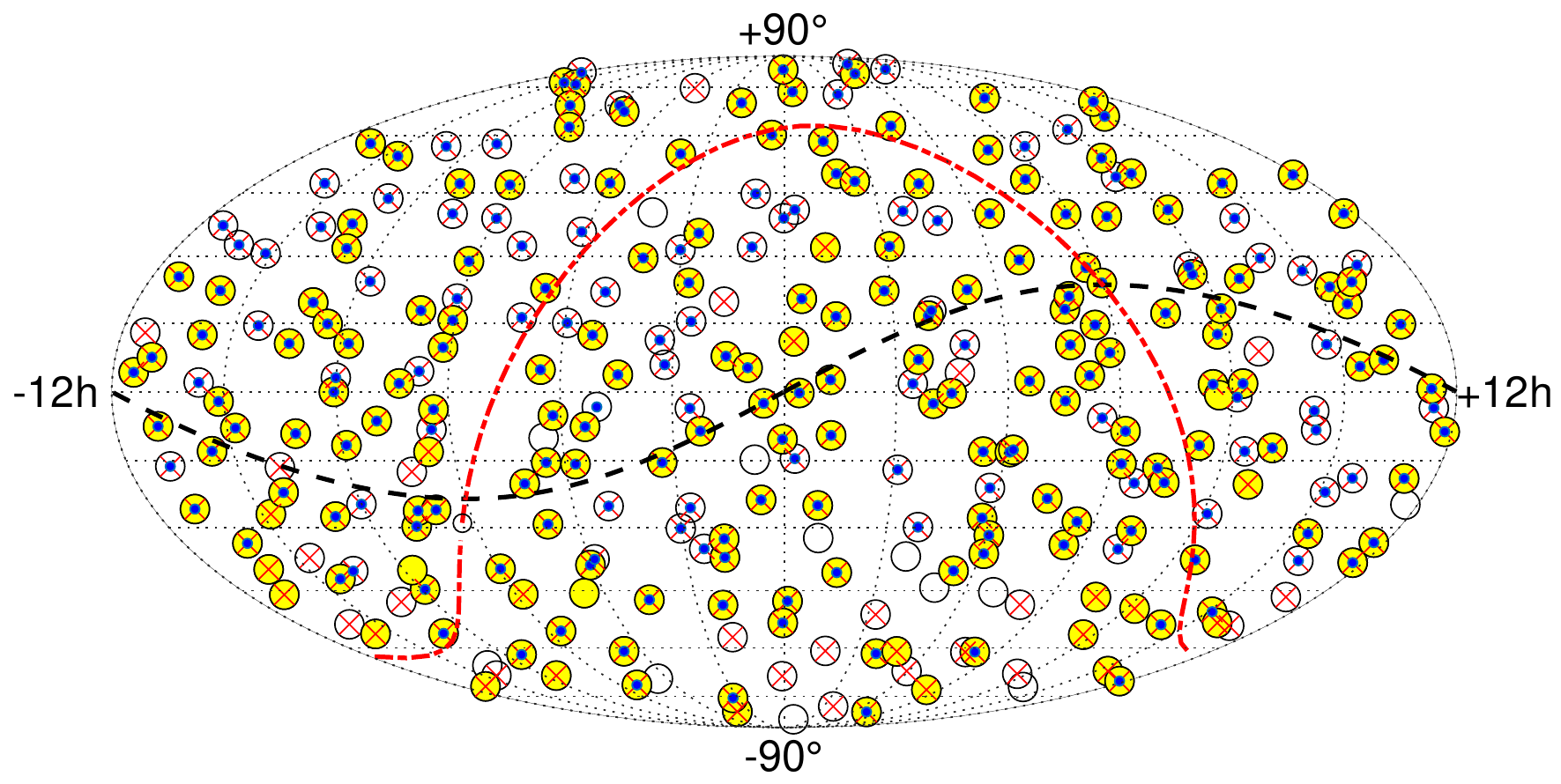}
\caption{Defining sources. The circles denote the 303 ICRF3-SX defining sources. The subgroup of 193 yellow circles depicts the ICRF3-K defining sources. Defining sources in \viecrfname{} and \usnocrfname{} are red crosses and blue dots, respectively.}\label{fig_defining}
\end{figure}

During the development of the ICRF3 a new set of sources observed at S/X-band was selected for defining the rotational alignment. This set of {\it defining} sources was based on three selection criteria in order to align the S/X-frame with its predecessor, the ICRF2 \citep{Fey2015}. These criteria were: (1) the overall sky distribution of the defining sources, (2) the position stability of the individual sources, and (3) the compactness of their structures \citep{Charlot20}. For the alignment of the K-band reference frame ICRF3-K, a subset of 193 sources out of the set of 303 ICRF3-S/X defining sources---based mainly on the number of available K-band observations---was used. In Fig.~\ref{fig_defining} we show the distribution of the ICRF3-SX defining sources and highlight the ICRF3-K defining sources with yellow color. In the solutions \viecrfname{} (red crosses) and \usnocrfname{} (blue dots) we take advantage of the additional observations gained after the ICRF3 release and choose the defining sources independently of the ICRF3-K ones. The current analysis of available sessions shows that there are no K-band observations of four ICRF3-S/X defining sources: 0044-846, 0855-716, 1448-648, 1935-692. This means, that 299 out of the 303 ICRF3-SX defining sources are observed in K-band (considering June 2022 to be the cutoff date for K-band observations).
In \viecrfname{} and \usnocrfname{} we apply different strategies for the selection of defining sources. 

At TU Wien, we first computed a K-CRF solution from VLBA sessions only. We found 12 AGN (0038-326,
0227-369,
0316-444,
0437-454,
0743-006,
1143-245,
1606-398,
1929-457,
1937-101,
2036-034,
2111+400,
2325-150) among the 303 ICRF3-SX defining sources whose angular separation in this VLBA-only K-CRF solution is greater than 0.5~mas from their ICRF3-SX position and those are dropped from the NNR condition in \viecrfname{}. All ICRF3-SX defining sources observed in the KS sessions only are kept in the NNR in \viecrfname{}.

In \usnocrfname{} the following sources were excluded from the defining set: 0700-465, 0742-562, 0809-493 and 0918-534 
since they show offsets of $0.5 - 1.5$ mas from their ICRF3-SX positions in recent USNO S/X solutions. An additional 41 sources, mostly in the deep south, were also excluded from the NNR condition because they had either very few or no observations. 

The rotation angles in Table~\ref{tab:souVSH} show that the incorporation of the deep south sources in the alignment condition makes the adjustment more robust and keeps the estimated K-CRF solution slightly closer to the a priori one.

\subsection{Ionospheric mapping function}
\label{sec_iono}

The global ionosphere maps provide the Vertical Total Electron Content (VTEC). The conversion from VTEC to the Slant Total Electron Content (STEC) at an elevation angle ($\epsilon$) of the VLBI observations at the telescope is done by the ionospheric mapping function (mf, $M$). In \viecrfname{} we apply the thin shell ionospheric mf introduced by \cite{Schaer1999} and recently discussed in detail by \cite{Petrov2022}:
\begin{equation}
M(\epsilon) = k\cdot \frac{1}{\sqrt{1-\Big(\frac{R_E}{R_E + H_i + \Delta H}\Big)^2 \cdot \cos^2\alpha\epsilon}} ,
\label{mf}
\end{equation}
where $k$ is a scaling factor, $R_E$ = 6371~km stands for the Earth's base radius, $H_i$ = 450~km is the height of the spherical single layer, $\Delta H$ represents an increment in the ionosphere height, and $\alpha$ is a correction factor to the elevation angle. In the default \viecrfname{} solution we apply: $k$ = 1, $\Delta H$ = 56.7~km, $\alpha$ = 0.9782 which is denoted as Modified Single-Layer Model (MSLM)\footnote{\href{http://ftp.aiub.unibe.ch/users/schaer/igsiono/doc/mslm.pdf}{http://ftp.aiub.unibe.ch/users/schaer/igsiono/doc/mslm.pdf}} mapping function and claimed to be the best fit with respect to the JPL extended slab model mapping function. This parameter setting is recommended e.g. by \cite{Feltens2018}, \cite{Wielgosz2021}, and references therein. The standard Single Layer Model (SLM) mapping function is achieved with the parameters: $k$ = 1, $\Delta H$ = 0~km, and $\alpha$ = 1. Following the discussion in \cite{Petrov2022}, we calculated two more solutions with different ionospheric mf parametrizations based on MSLM with different values of $\Delta H$ and $k$ (i.e., iono3 and iono4) as summarized in Table~\ref{tab:iono}.

\begin{figure}[h]%
\centering
\includegraphics[width=0.5\textwidth]{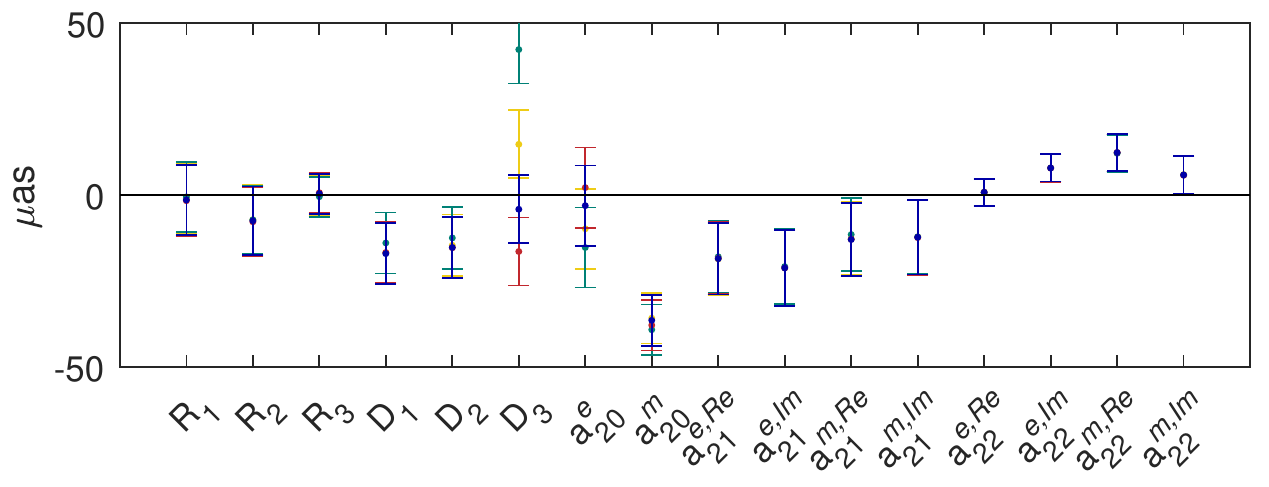}
\caption{Vector spherical harmonics of K-CRF solutions computed with ionospheric mf MSLM (blue, \viecrfname{}), SLM (red), iono3 (yellow), and iono4 (green) w.r.t. ICRF3-SX.}\label{fig_vsh}
\end{figure}

\begin{figure}[h]%
\centering
\includegraphics[width=0.5\textwidth]{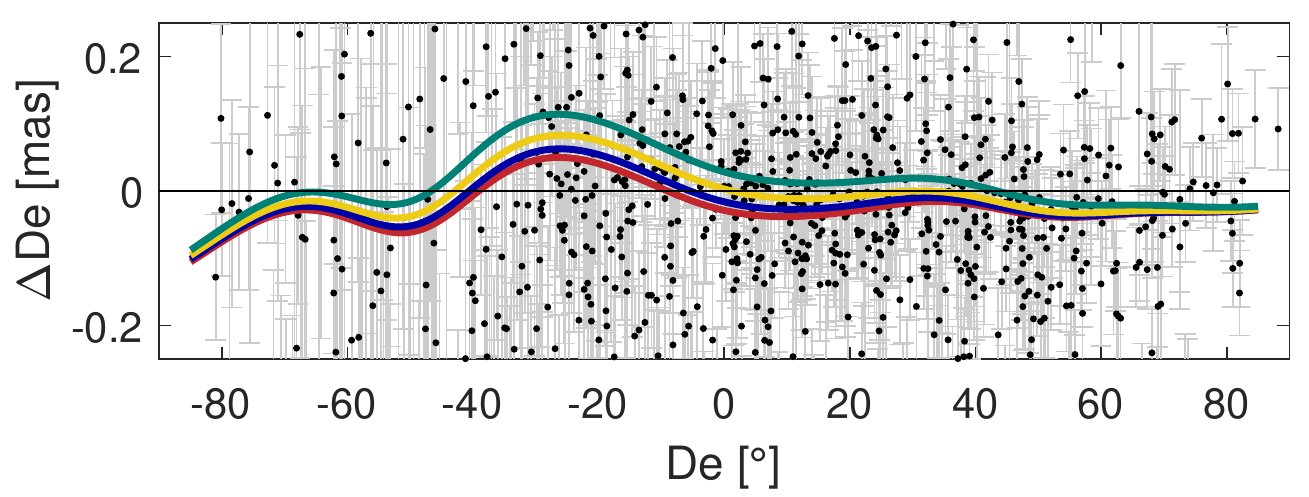}
\caption{Smoothed differences in declination from K-CRF solutions computed with ionospheric mf MSLM (blue, \viecrfname{}), SLM (red), iono3 (yellow), and iono4 (green) w.r.t. ICRF3-SX. The black dots are differences in the declination of individual sources in \viecrfname{} w.r.t. ICRF3-SX and their formal errors (in grey).}\label{fig_de}
\end{figure}

In order to quantify the effect of the modified ionospheric mapping function on the K-CRF solution, we calculated VSH for each solution w.r.t. ICRF-SX (Fig.~\ref{fig_vsh}). Changes in the three mf parameters ($k$, $\Delta H$, $\alpha$) influence the terms $D_3$ and $a_{2,0}^{e}$, which are sensitive to the equatorial bulge and north-south asymmetries, as mentioned earlier. The best fit to the ICRF3-SX is achieved with the MSLM mapping function applied in \viecrfname{} where these parameters are negligibly small ($ -4 \pm 10$ \textmu as and $ -3 \pm 12$ \textmu as, respectively). On the other hand, in iono4 (where a scale factor $k$ = 0.85 is applied to MSLM), the difference w.r.t. ICRF3-SX in $D_3$ and $a_{2,0}^{e}$ increases to $42 \pm 10$ \textmu as  and  $-15 \pm 12$ \textmu as, respectively. In Fig.~\ref{fig_de} we plot the differences in declination between the four discussed solutions w.r.t. ICRF3-SX over declination for individual sources. The smoothed curves are computed as moving averages with a Gaussian kernel and plotted with color coding identical to Fig.~\ref{fig_vsh}. The positive systematic difference in the declination estimates w.r.t. ICRF3-SX, appearing approximately between -40$^{\circ}$ and -10$^{\circ}$ declination, reaches its maximum of 63 \textmu as for -26$^{\circ}$ declination in \viecrfname{} with applied MSLM mapping function (blue curve).

\begin{table}
\caption{Parameters of the ionospheric mapping function and the resulting VSH parameters $D_3$ and $a_{2,0}^{e}$. }
\label{tab:iono}       
\begin{tabular}{l|r|r|r|r|r}
\hline\noalign{\smallskip}
    & $k$ & $\Delta H$ & $\alpha$ & $D_3$ & $a_{2,0}^{e}$  \\
        & [-] & [km] & [-] & [\textmu as] & [\textmu as]  \\
\noalign{\smallskip}
\hline 
MSLM &  1	&  56.7 &   0.9782 & $ -4 \pm 10$ & $ -3 \pm 12$\\ 
SLM &  1	&  0    &   1 & $-17 \pm 10$ & $ +2 \pm 12 $\\ 
iono3 &  1	&  150.0  &   0.9782 & $ 15 \pm 10  $  &   $ -10 \pm 12 $ \\ 
iono4 & 0.85 & 56.7 & 	0.9782 & $42 \pm 10 $   &    $-15 \pm 12$\\
\noalign{\smallskip}\hline
\end{tabular}
\end{table}

\subsection{Systematic in elevation angles} 
\label{sec_syselev}

\begin{figure}[h]%
\centering
\includegraphics[width=0.5\textwidth]{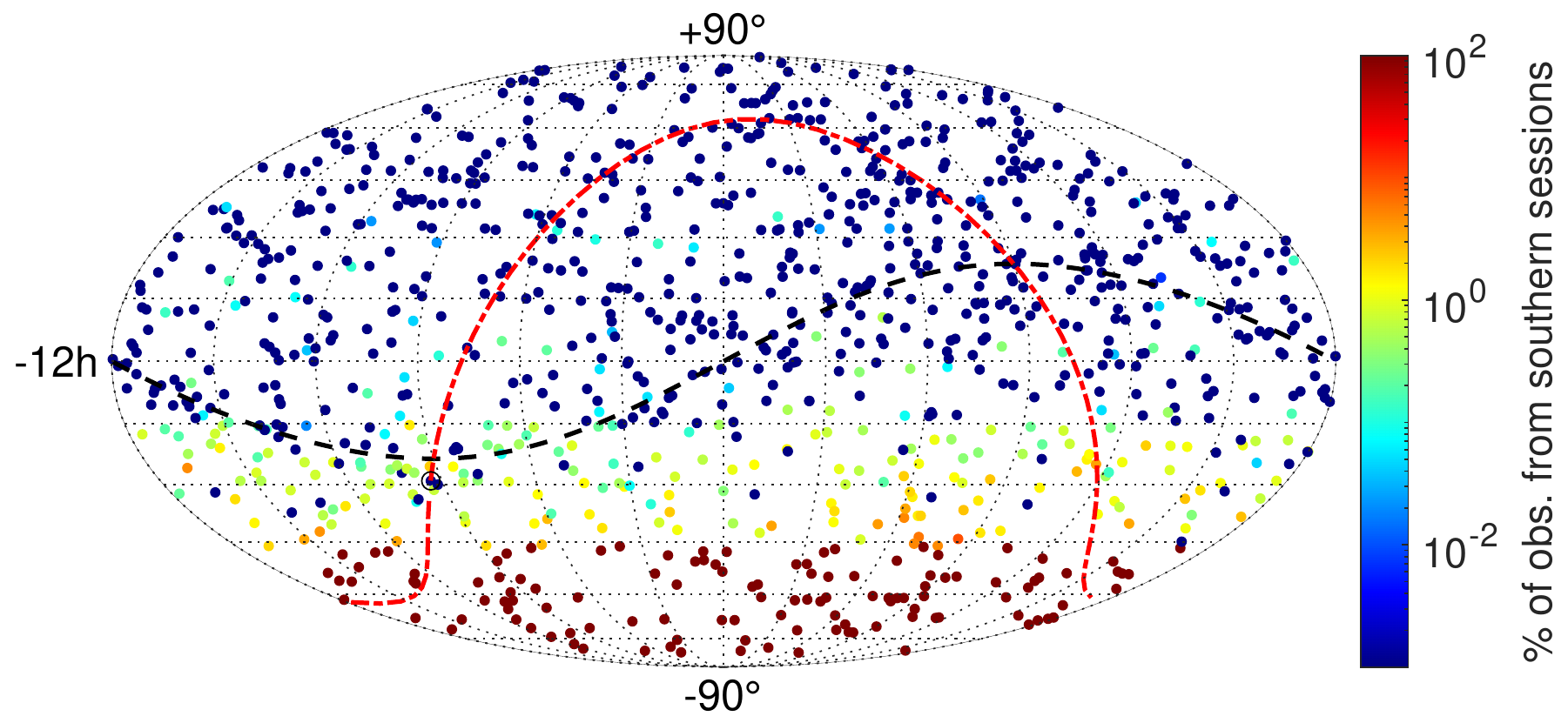}
\caption{Percentage of observations from KS sessions among the total number of K-CRF observations for individual sources in \viecrfname{}.} \label{fig_numobs_DSwrtVLBA}
\end{figure}

\begin{figure}[h]%
\centering
\includegraphics[width=0.5\textwidth]{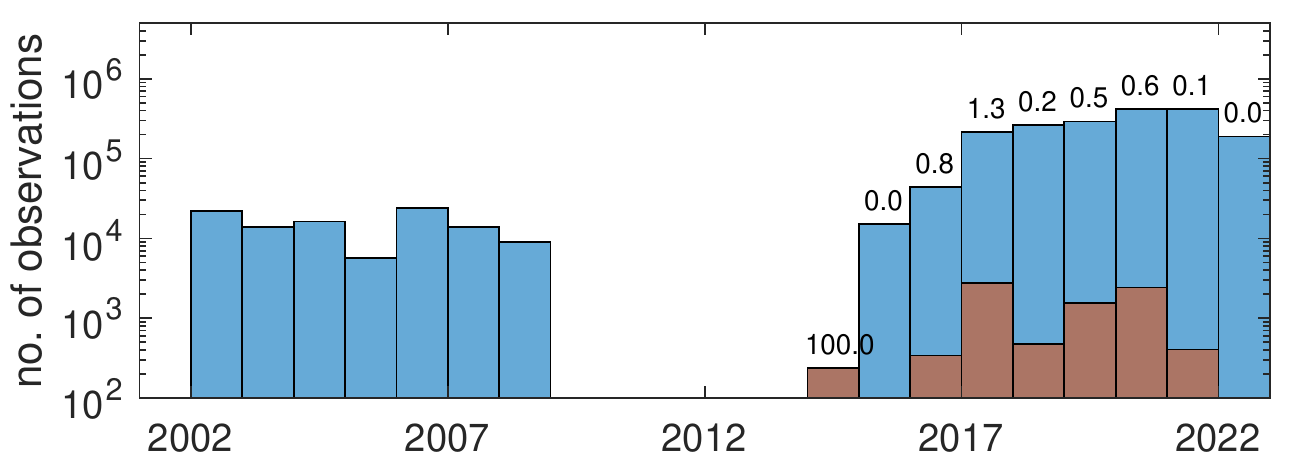}
\caption{The yearly distribution of the K-CRF observations used in our solutions. The numbers above the columns show the percentage of observations from southern sessions (KS, brown columns) w.r.t. the total number of observations (blue columns) during the individual year.}\label{fig_numobs_histogram}
\end{figure}

Along with the ionospheric effects, the K-CRF suffers from an asymmetric observing network geometry with 99\% of the data being from the all-northern VLBA. In Fig.~\ref{fig_numobs_DSwrtVLBA} the percentage of observations from southern KS sessions for individual sources in \viecrfname{} is shown. The logarithmic color scale highlights the fact that the number of observations to the sources with declination higher than $-45^{\circ}$ builds only a tiny fraction of the total number of observations, although the mutual sky visibility between southern antennas HartRAO and Hobart allows observing sources up to approximately $30^{\circ}$ declination. The mean percentage of observations from KS sessions for sources with declination between $-15^{\circ}$ and $-45^{\circ}$ (area with mainly yellow and green colors in Fig.~\ref{fig_numobs_DSwrtVLBA}) is 0.96\%. The total number of K-CRF observations (blue color) and the number of observations from the KS (brown color) during individual years is plotted in Fig.~\ref{fig_numobs_histogram}. The numbers above the columns give the percentage of observations from KS w.r.t. the total number of observations within the individual year.

\begin{figure}[h]%
\centering
\includegraphics[width=0.5\textwidth]{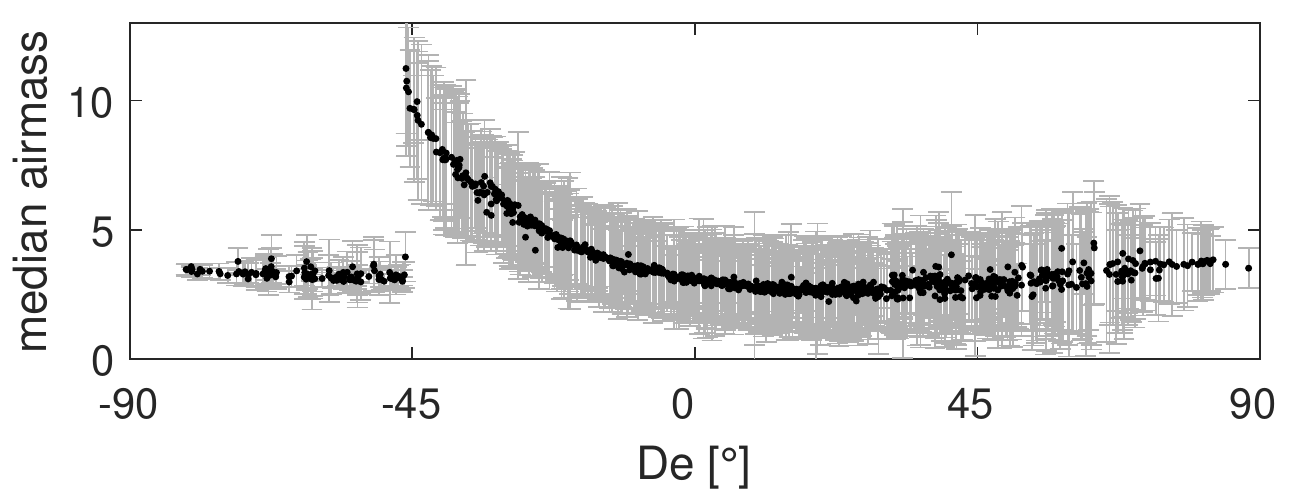}
\caption{Median airmass for individual sources computed over all their observations in \viecrfname{}.}\label{fig_airmass}
\end{figure}

In order to explore the resultant elevation-dependent effects, we characterize the distribution of elevation angles at which the sources were observed. These distributions vary due to both the geometry of the VLBA network and the fact that we observe each source over a range of hour angles. First, we define a parameter called airmass in order to quantify the approximate total pathlength through the troposphere for each source---with the maximum at lower elevation angles. It is computed for each observation from the whole data set with the simplifying assumption of a flat slab atmosphere (ignoring the curvature of the atmosphere over a spherical Earth):
\begin{equation}
\mathrm{airmass} = \frac{1}{\sin(\epsilon_1)} + \frac{1}{\sin(\epsilon_2)} ,
\label{eq_airmass}
\end{equation}
where $\epsilon$ is the elevation angle of the source at telescopes 1 and 2 of the baseline. Next, we compute the median value over the individual observations for each source and plot it with respect to the declination (Fig.~\ref{fig_airmass}) with the errors (in grey) obtained as standard deviations computed over the individual airmass values for the particular source. The systematic increase of the airmass parameter from  0$^{\circ}$ to $-45^{\circ}$ declination can lead to an overestimation of the optimal data weights for VLBA observations in this declination range when the larger noise of observations conducted at low elevation angles is not considered. 
To partly account for the overweighting of the low elevation scans (which observe low declination sources in the mentioned area), elevation-dependent weighting \citep[Eq.~(\ref{eq_diag_elevdep});][] {Gipson08} in \viecrfname{} is applied. In the diagonal covariance matrix the measurement noise $\sigma^2_{m}$ is increased by the squared elevation-dependent noise terms for telescopes 1 and 2: 
\begin{equation}
\sigma^2_{obs} = \sigma^2_{m} + \bigg(\frac{6~\texttt{ps}}{\sin(\varepsilon_1)}\bigg)^2  + \bigg(\frac{6~\texttt{ps}}{\sin(\varepsilon_2)}\bigg)^2 .
\label{eq_diag_elevdep}
\end{equation}
Hence, sources between 0$^{\circ}$ to $-45^{\circ}$ declination obtain a lower weight in the least squares adjustment and the resulting distortion of the celestial reference frame is damped.  For example, an observation conducted with two VLBA antennas at the elevation angles of 15$^{\circ}$ has an airmass value of 8 (Eq.~(\ref{eq_airmass})) which corresponds in our data set to a source with a declination of about $-40^{\circ}$ (Fig.~\ref{fig_airmass}). The additional noise added to the $\sigma^2_{m}$ of this observation in quadrature is 33 ps (Eq.~(\ref{eq_diag_elevdep})) which decreases its weight in the solution.

\section{Conclusion}
Recent K-CRF solutions computed at TU Wien (\viecrfname{})\footnote{\href{https://vlbi.at/data/analysis/ggrf/crf_vie2022b_k.txt}{https://vlbi.at/data/analysis/ggrf/crf_vie2022b_k.txt}} and USNO (\usnocrfname{}) from single-frequency band VLBI observations (24~GHz) until June 2022 were assessed. The vector spherical harmonics were computed w.r.t. ICRF3-SX after eliminating four AGN as outliers. In VIE-K-2022b, all rotation values are lower than 8~\textmu as and have significance at the level of their formal errors or less. With a single exception, all dipole and quadrupole terms are within  20~\textmu as with a marginal significance of two times the formal error as maximum. The only quadrupole term above this limit is $a_{2,0}^{m}$  ($-36 \pm 7$ \textmu as). We discussed two major challenges which limit the accuracy of the current K-band VLBI solutions: external ionospheric corrections and the non-uniform observing network geometry---especially the lack of observations in the deep south. We show that the choice of ionospheric mapping function parameters influences the dipole, $D_3$, and quadrupole terms $a_{2,0}^{e}$. Because 99\% of the data is observed with the all-northern VLBA  sources between 0$^{\circ}$ and $-45^{\circ}$ declination have a monotonic decrease in median elevation angle of observation making our solution vulnerable to atmospheric mis-modeling. We reduced sensitivity of the \viecrfname{} solution to the effect of this observing geometry bias by computing elevation-dependent weighting to downweight low elevation observations.
Future work will focus on improving the geometry of the K-band observing network, improving the modeling of atmospheric effects, and improving solution weighting schemes.

\backmatter

\section{Declarations}
\bmhead{Ethics approval and consent to participate}
Not applicable.
\bmhead{Consent for publication}
Not applicable.
\bmhead{Competing interests}
There are no relevant financial or non-financial competing interests to report.
\bmhead{Funding}
We acknowledge our respective sponsors: SARAO/HartRAO is a facility of the National Research Foundation (NRF) of South Africa. Portions of this work were done at the Jet Propulsion Laboratory, California Institute of Technology under contract with NASA (contract no. 80NM0018D0004). Portions of this work were sponsored by the Radio Optical Reference Frame Division of the U.S. Naval Observatory. This work supports USNO’s ongoing research into the celestial reference frame and geodesy.  
\bmhead{Authors' contributions}
HK wrote the manuscript, analyzed the VLBI data and created the VIE-K solutions. DG prepared the vgosDB databases and computed the USNO-K solution. AdW is the PI of the VLBI K-band group and the leader by planning of the VLBI K-band observations. CJ proposed the concept of the paper and contributed to the analysis of data. All authors contributed to regular discussions and interpretations of results. They read and commented the final paper.
\bmhead{Acknowledgments}
The authors appreciate comments provided by three anonymous reviewers. HK thanks Leonid Petrov (NASA GSFC) for fruitful discussions about single band astrometry. The authors gratefully acknowledge the use of the VLBA under the USNO’s time allocation. 

\bibliography{reference_krasna} 



\end{document}